\documentclass[10pt,journal,final,twoside]{IEEEtran}

\ifCLASSINFOpdf
  \usepackage[pdftex]{graphicx}
  \graphicspath{{figures/}}
  \DeclareGraphicsExtensions{.pdf,.png}
\else
  \usepackage[dvips]{graphicx}
  \graphicspath{{figures/}}
  \DeclareGraphicsExtensions{.pdf,.png}
\fi

\usepackage{amsmath}
\usepackage{amsfonts}

\usepackage{multirow}
\usepackage[caption=false,font=footnotesize]{subfig}

\usepackage{breakurl}

\usepackage{booktabs} 
\usepackage[inline]{enumitem}
\usepackage[linesnumbered ]{algorithm2e} 
\usepackage[utf8]{inputenc}
\usepackage{comment}

\begin{document}

\title{RouteNet: Leveraging Graph Neural Networks for network modeling and optimization in SDN}

\author{Krzysztof~Rusek,
        Jos\'e~Su\'arez-Varela,
        Paul~Almasan,
        Pere~Barlet-Ros
        and~Albert~Cabellos-Aparicio
        \\ \vspace{0.4cm}
        \small{\textbf{NOTE:} This is an extended version of a paper published in IEEE JSAC. Please use the following reference to cite this work:\\ K. Rusek, J. Suárez-Varela, P. Almasan, P. Barlet-Ros and A. Cabellos-Aparicio, "RouteNet: Leveraging Graph Neural Networks for network modeling and optimization in SDN," in IEEE Journal on Selected Areas in Communications, doi: 10.1109/JSAC.2020.3000405.}
        \vspace{-1cm}
\thanks{Krzysztof~Rusek is with the Department of Telecommunications, AGH University of Science and Technology, Krakow, Poland, and with the Barcelona Neural Networking Center, Universitat Polit\`ecnica de
Catalunya, Barcelona, Spain. (e-mail: krusek@agh.edu.pl).}
\thanks{Jos\'e~Su\'arez-Varela, Paul~Almasan, Pere~Barlet-Ros~and~Albert~Cabellos-Aparicio are with the Barcelona Neural Networking Center, Universitat Polit\`ecnica de
Catalunya, Barcelona, Spain (e-mail: \{jsuarezv,almasan,pbarlet,acabello\}@ac.upc.edu).}
\thanks{This work was supported by the Polish Ministry of Science and Higher Education with the subvention funds of the Faculty of Computer Science, Electronics and Telecommunications of AGH University, the Spanish MINECO under contract TEC2017-90034-C2-1-R (ALLIANCE), the Catalan Institution for Research and Advanced Studies (ICREA) and the FI-AGAUR grant by the Catalan Government. The research was also supported in part by PL-Grid Infrastructure.}
}
\vspace{0.2cm}

\markboth{IEEE JOURNAL ON SELECTED AREAS IN COMMUNICATIONS,~Vol.~X, No.~X,~MONTH,~YYYY}{Rusek and Su\'arez-Varela \MakeLowercase{\textit{et al.}}}

\IEEEoverridecommandlockouts
\IEEEpubid{\makebox[\columnwidth]{10.1109/JSAC.2020.3000405~\copyright2020 IEEE \hfill} \hspace{\columnsep}\makebox[\columnwidth]{ }}

\maketitle

\begin{abstract}
Network modeling is a key enabler to achieve efficient network operation in future self-driving Software-Defined Networks. However, we still lack functional network models able to produce accurate predictions of Key Performance Indicators (KPI) such as delay, jitter or loss at limited cost. In this paper we
propose RouteNet, a novel network model based on Graph Neural Network (GNN) that is able to understand the complex relationship between topology,
routing, and input traffic to produce accurate estimates of the
per-source/destination per-packet delay distribution and loss. RouteNet leverages the ability of GNNs to learn and model graph-structured information
and as a result, our model is able to generalize over arbitrary
topologies, routing schemes and traffic intensity. In our evaluation, we show that RouteNet is able to predict accurately the delay distribution (mean delay and jitter) and loss even in topologies, routing and traffic unseen in the training (worst case MRE=15.4\%). Also, we present several use cases where we leverage the KPI predictions of our GNN model to achieve efficient routing optimization and network planning.
\end{abstract}


\IEEEpeerreviewmaketitle

\section{Introduction}

Network modeling is a fundamental component to achieve efficient network optimization with special attention on future self-driving networks~\cite{kdn}. In the context of Software-Defined Networks, networking tasks are orchestrated from a centralized control plane, which may leverage a global picture of the network state in order to operate networks efficiently and dynamically adapt to changes in the network. To this end, network administrators typically define a target policy that may include some optimization objectives (e.g., minimize end-to-end latency) and constraints (e.g., security policy). Then, SDN controllers are tasked to find some changes in the network configuration (e.g., routing) to accomplish the optimization objectives set by administrators. This is typically achieved by combining two main elements: $(i)$ a network model, and $(ii)$ an optimization algorithm. In this well-known optimization architecture, the network model is tasked to predict the resulting performance (e.g, delay, packet loss) for specific configurations, and the optimization algorithm iteratively explores different configurations until it finds one that meets the optimization goals.

One fundamental issue of network optimization solutions is that they can only optimize based on the performance metrics provided by the network model. Thus, in order to optimize Key Performance Indicators (KPI) such as delay or packet loss in networks, it is essential a network model able to understand how these performance indicators are related to the network state metrics collected from the data plane, which often can provide only timely statistics of traffic volume (e.g., traffic matrix) in real-world deployments. In this context, much effort has been devoted in the past to build network models able to predict performance metrics, however nowadays we still lack functional models providing accurate predictions of relevant KPI like delay, jitter or packet loss. Analytic models, mainly based on Queuing Theory \cite{queuingModels}, assume some non-realistic properties of networks (e.g., traffic with Poisson distribution, probabilistic routing) and, as a result, they are not accurate to produce KPI predictions in large-scale networks with realistic configurations such as multi-hop routing~\cite{experienceDriven}. Conversely, packet-level network simulators showed to be very accurate for this purpose, but their high computational cost makes it unfeasible to leverage them to operate networks in short time scales.

In this context, Deep Learning~\cite{deepLearning} seems to be a well-suited alternative to develop a new breed of network models that can be both accurate and lightweight. Relevant research efforts are being devoted to apply neural networks to model computer networks~\cite{wangMachineLearning} and using such models for network optimization \cite{IntelligentRouting,learningRouting,experienceDriven}. Existing proposals \cite{deepQ,mestresModeling} typically used well-known Neural Networks (NN) architectures like fully-connected Neural Networks, Convolutional Neural Networks, Recurrent Neural Networks or Variational Auto-Encoders. However, computer networks are fundamentally represented as graphs, and such types of NN are not designed to learn graph-structured information. As a result, the models trained result in limited accuracy and are unable to generalize in terms of topologies or routing configurations.

In this paper we present RouteNet, a novel network model based on Graph Neural Networks (GNN)~\cite{graphNetworks}. Our model is able to understand the complex relationship between topology, routing, and input traffic to accurately estimate the distribution of the per-packet delay and loss ratio on every source-destination pair. GNNs are tailored to achieve relational reasoning and combinatorial generalization over information structured as graphs~\cite{relationalInductiveBias} and as a result our model is able to generalize over arbitrary topologies, routing schemes and variable traffic intensity. In particular, RouteNet captures \emph{meaningfully} traffic routing over network topologies. This is achieved by modeling the relationships of the links in topologies with the source-destination paths resulting from the routing schemes and the traffic flowing through them. 
One main contribution of the RouteNet architecture compared to
other GNN based models~\cite{Geyer2019} is the 
representation of paths as ordered sequences of links.
This makes RouteNet a new GNN architecture designed especially for 
computer network control and management. 

An earlier version of this paper was presented at~\cite{routenetsosr2019}.
In that version, two different models were used to predict the per-path mean delay and jitter. In this paper we present an extended RouteNet model inspired by Generalized Linear Models that directly estimates the per-packet distribution of the delay on each path. This enables to use a single model to predict any metric associated to end-to-end per-packet delay (e.g., mean delay, jitter). Additionally, in this paper we adapted RouteNet to make also predictions of the per-source/destination packet loss ratio.

We evaluated the accuracy of our GNN model with a dataset generated using a packet-level simulator (Omnet++~\cite{omnet}), and this resulted in high estimation accuracy of delay, jitter, and loss when testing it against topologies, routing and traffic not seen during training. More importantly, we verify that our model is able to generalize and, for instance, when training the model with samples of 14-node, 24-node and 50-node topologies the model is able to provide accurate estimates in a never-seen 17-node network (MRE=15.4\% in the worst case).

Finally, and in order to showcase the potential of our GNN model we present a series of use cases applicable to a SDN architecture. In contrast to the use cases presented in~\cite{routenetsosr2019}, in this paper we include network scenarios that leverage also the new RouteNet model that predicts the packet loss to perform a joint optimization of mean delay, jitter, and loss. We first show that RouteNet can be used to optimize the routing configuration in QoS-aware scenarios with delay, jitter and loss requirements, and benchmark it against traditional utilization-aware models (e.g., OSPF) and the optimal solution using a packet-level simulator. Also, we leverage the predictions of RouteNet in a network planning use case to select the optimal link placement.

We summarize below the main contributions of this paper compared to the previous one and the state-of-the-art:
\begin{itemize}
	\item Probabilistic modeling inspired by Generalized Linear Models 
	\item Adaptation to predictions of the per-source/destination packet loss ratio
	\item Residual connection to facilitate the training (like in ResNet~\cite{resnet})
	\item Computation cost improvement ($\sim$10$\times$ faster)
	\item Additional input features (support for arbitrary link capacities)
	\item New network optimization use cases incorporating packet loss requirements.
	\item More diverse and larger network topologies (with 14, 17, 24 and 50 nodes)
\end{itemize}

\section{SDN-based Modeling and Optimization Scenario}

\begin{figure}[!t]
	\centering
	\includegraphics[width=1.0\linewidth]{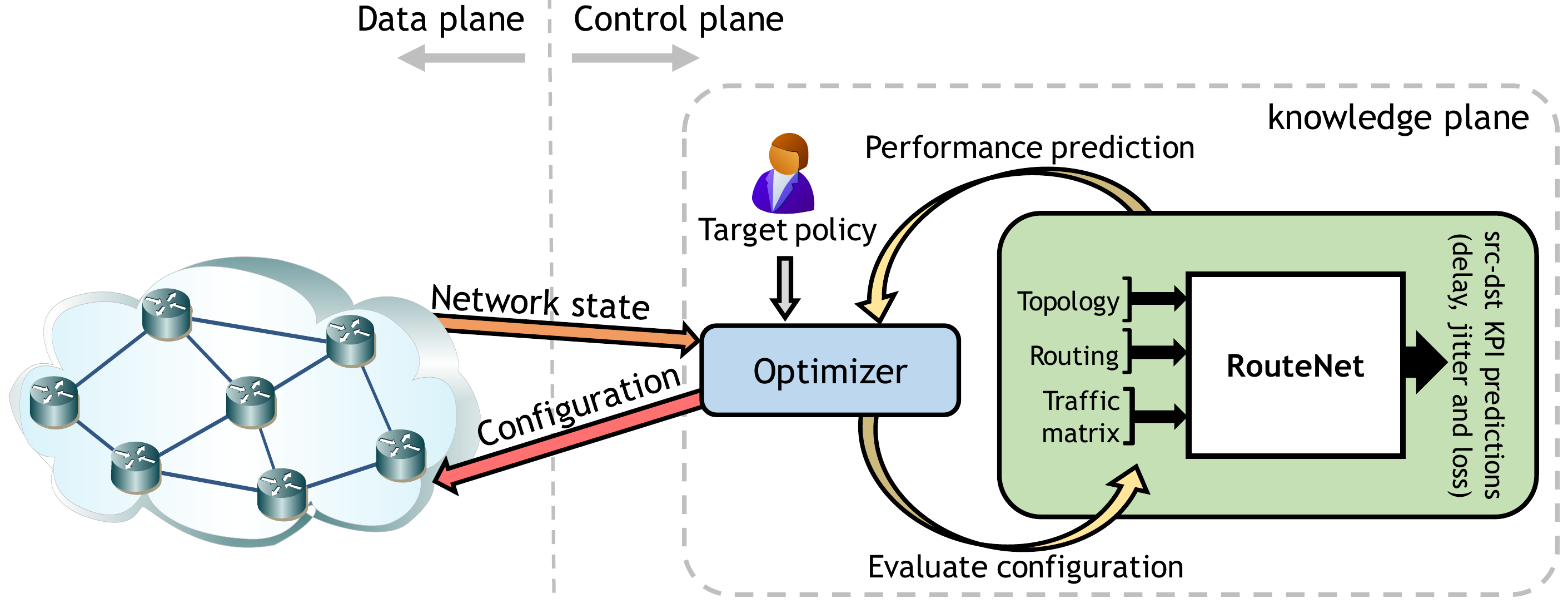}
	\caption{Architecture for network optimization in SDN}
	\label{fig:scenario}
\end{figure}

Network modeling enables the control plane to further exploit the potential of SDN to perform fine-grained management. This permits to evaluate the resulting performance of what-if scenarios without the necessity to modify the state of the data plane. It may be profitable for a number of network control and management applications such as optimization, planning or fast failure recovery. For instance, in Fig.~\ref{fig:scenario} we show an architecture for network optimization within the context of the \textit{knowledge-Defined Networking} (KDN) paradigm~\cite{kdn}. In this case, we assume that the control plane receives timely updates of the network state (e.g., traffic matrix, delay measurements). This can be achieved by means of ``conventional'' SDN-based measurement techniques (e.g., OpenFlow~\cite{openflow}, OpenSketch~\cite{opensketch}) or more novel telemetry proposals such as INT for P4 \cite{int-p4} or iOAM~\cite{ioam}. Likewise, in the knowledge plane there is an optimizer whose behavior is defined by a given target policy. This policy, in line with intent-based networking, may be defined by a declarative language such as NEMO \cite{nemo} and finally 
being translated to a (multi-objective) network optimization problem. At this point, an accurate network model can play a crucial role in the optimization process 
by leveraging it to run optimization algorithms (e.g., hill-climbing) that iteratively explore the performance of candidate solutions in order to find the optimal configuration. We intentionally leave out of the scope of this architecture the training phase.

To be successful in scenarios like the one proposed above, the network model should meet two main requirements: \newline $(i)$ accurate performance prediction, and $(ii)$ low computational cost to enable network operation in short time scales. Moreover, it is essential for optimizers to have enough flexibility to predict the resulting performance after changing the network configuration, vary the input traffic, or modify the topology (e.g., link failure). To this end, we rely on the capability of Graph Neural Networks (GNN) to
efficiently operate and generalize over graph-structured data. RouteNet, the GNN-based model proposed in this paper, is able to propagate any routing scheme throughout a network topology and abstract meaningful information of the current network state to produce relevant performance 
estimates. More in detail, RouteNet (Fig.~\ref{fig:scenario}) takes as input $(i)$ a given topology, $(ii)$ a source-destination routing scheme (i.e., list of end-to-end paths) and $(iii)$ a traffic matrix (defined as the bandwidth between each node pair in the network), and produces as output performance metrics according to the current network state (per-path mean delay, jitter, and packet loss). To achieve it, RouteNet uses fixed-dimension vectors that encode information about the state of paths and links and propagate the information among them according to the input topology and the routing scheme.

\section{Network Modeling with GNN}
\label{sec:modeling-gnn}

\subsection{Notation}
A computer network can be represented by a set of links $\mathcal{N}=\{l_i| i \in (1,\ldots,n_l)\}$,
and the routing scheme in the network by a set of paths $\mathcal{R}=\{p_k| k \in (1,\ldots,n_p) \}$. Each path is defined as a sequence of links $p_k=(l_{k(1)},\ldots, l_{k(|p_k|)})$, where $k(i)$ is the index of the $i$-th link in the path $k$. The properties (features) of both links and paths are denoted by $\mathbf x_{l_i}$ and $\mathbf x_{p_i}$. 
Measurable KPIs are modeled as random variables  $W_{i}$ and $L_{i}$, where the former is the end-to-end delay and the later is the total number of packet drops during a period of time for every source-destination pair in the network.

\subsection{Background on Graph Neural Networks}
Graph neural network is an artificial neural architecture designed for graph-structured data, where the nodes, edges, and the whole graph can have associated feature vectors. The most important property of GNN is that it preserves the basic topological relations between node adjacencies (graph isomorphism), so it 
is well suited to be used for different topologies without retraining.

Multiple GNN architectures have been proposed at the time of this writing. Recently, Message Passing Neural Network (MPNN) was proposed as a general family of 
GNN architectures~\cite{MPNN}. Most of the 
existing GNN models 
can be described as special cases of the MPNN framework.

The main assumption of MPNN is that the information related to nodes, edges or the whole graph can be encoded in 
fixed-dimension vectors, also called embeddings. 
The forward pass in MPNN is 
a combination of
three simple functions: $(i)$ Message, $(ii)$ Update, and $(iii)$ Readout. 
The Message function takes as input node/edge embeddings, and it outputs an information vector (the message) to be sent to all the neighbors in the graph. The Update function collects (sums) messages from all the neighbors in the graph and updates the embeddings of the nodes/edges. This message exchange is repeated $T$ times, and 
finally the Readout function takes the resulting nodes/edges embeddings to produce the output of the GNN model. 
We built upon this concept to construct RouteNet -- a message passing architecture 
specifically tailored to produce accurate performance estimates in computer networks.

\subsection{Message Passing Architecture Of RouteNet}

RouteNet handles variable-size input network topologies and arbitrary source-destination routing schemes, and produces as output end-to-end performance predictions. 
The main assumption behind RouteNet is that information at the path level (e.g., end-to-end metrics such as delays or packet loss) and the link level (e.g., link delay, packet loss rate, link utilization) can be encoded in learnable vectors of real numbers (path and link state vectors respectively).
Note that the path abstraction may not necessarily correspond to a physical path. It could be a generic end-to-end traffic flow. For instance, an MPLS tunnel. Based on this assumption, RouteNet is built upon the following principles:
\begin{enumerate}[label=\arabic*)]
	\item The state of a path depends on the state of all the links that lie on the path.
	\item The state of a link depends on the state of all the paths that traverse the link.
\end{enumerate}
In a more formal description, let the state of a link be denoted by $\mathbf{h}_{l_i}$, which is an unknown hidden vector.
Similarly, the state of a path is defined by $\mathbf{h}_{p_i}$.
These principles can be mathematically formulated with the following expressions:
\begin{align}
&\mathbf{h}_{l_i} = f(\mathbf{h}_{p_1},\ldots, \mathbf{h}_{p_j}),  \quad l_i \in p_k,\space k=1, \ldots, j \label{eq:hl}\\ 
&\mathbf{h}_{p_k} = g(\mathbf{h}_{l_{k(1)}},\ldots, \mathbf{h}_{l_{k(|p_k|)}}) \label{eq:hp}
\end{align}
where $f$ and $g$ are some unknown functions. 
It is well-known that neural networks can work as universal function approximators.
However, a direct approximation of functions $f$ and $g$ is not possible in this case given that:
$(i)$ Equations~\eqref{eq:hl} and~\eqref{eq:hp} define an implicit function (a nonlinear system of equations with the states being hidden variables),
$(ii)$ these functions depend on the input routing scheme, and
$(iii)$ the dimensionality of each function is very large. This would require a vast set of training samples.

RouteNet represents a GNN architecture that learns efficiently $f$ and $g$.
It is invariant to the topology and routing scheme and makes the neural function approximation feasible. Algorithm~\ref{alg:mpnn} describes the forward propagation (and the internal architecture) of RouteNet. In this process, this GNN model receives as input the initial path and link features $\mathbf x_p$, $\mathbf x_l$ and the routing description $\mathcal{R}$, and outputs inferred per-path metrics ($\hat{\mathbf y}_p$). Note that we simplified the notation by dropping sub-indexes of paths and links.

\begin{algorithm}[t]
	\SetAlgoLined
	\KwIn{$\mathbf x_p$, $\mathbf x_l$,$\mathcal{R}$} \label{lin:in}
	\KwOut{$\mathbf h_p^{T}$, $\mathbf h_l^{T}$, $\hat{\mathbf y}_p$}
	\lForEach{$p \in \mathcal{R}$}{$\mathbf h_p^0 \leftarrow [\mathbf x_p,0\ldots, 0]$}\label{lin:hp0}
	\lForEach{$l \in \mathcal{N}$}{$\mathbf h_l^0 \leftarrow [\mathbf x_l,0\ldots, 0]$}\label{lin:hl0}
	\For{$t=0$ to $T-1$}{\label{lin:forT}
		\ForEach{$p \in \mathcal{R}$}{
			\ForEach{$l \in p$}{\label{lin:mp}
				$\mathbf h_p^{t} \leftarrow RNN_t(\mathbf h_p^t,\mathbf h_l^t)$ \label{lin:hp} \\
				$\tilde{\mathbf m}_{p,l}^{t+1}\leftarrow \mathbf h_p^{t} $
			}
			$\mathbf h_p^{t+1} \leftarrow \mathbf h_p^{t}$ \label{lin:up}
		}
		\ForEach{$l \in \mathcal{N}$}{
			$ \mathbf h_l^{t+1} \leftarrow U_t\left(\mathbf h_l^t,\sum_{p:k \in p} \tilde{\mathbf m}_{p,k}^{t+1}\right)$ \label{lin:mlul}
		}
	}\label{lin:mp-end}
	$\hat{ y}_p \leftarrow F_p(\mathbf h_p)$\label{lin:read}
	\caption{Internal architecture of RouteNet. Complexity in number of nodes: $\sim O(n^2\log(n)), < O(n^3)$~\cite{routenetsosr2019}. }
	\label{alg:mpnn}
\end{algorithm}

RouteNet's architecture enables dealing with the circular dependencies described in equations \eqref{eq:hl} and \eqref{eq:hp}, and supporting arbitrary routing schemes (which are inherently represented within the architecture).
In order to address the circular dependencies, RouteNet repeats the same message passing operations over the links' and paths' state vectors $T$ times (loop from line~\ref{lin:forT}). These steps represent the convergence process to the fixed point of a function from the initial states $\mathbf{h}_{p}^{0}$ and $\mathbf{h}_{l}^{0}$.

Regarding the issue of routing invariance (more generically known as topology invariance in the context of graph-related problems), RouteNet requires the use of a structure able to represent graphs of different topologies of variable size. In our case, we aim at representing different routing schemes in a uniform way. One state-of-the-art solution for this problem~\cite{rusek2018message} proposes using neural message passing architectures that combine both: a representation of the topology as a graph, and vectors to encode the link states. In this context, RouteNet can be interpreted as an extension of a vanilla message passing neural network that is specifically suited to represent the dependencies among links and paths given a routing scheme (Equations \eqref{eq:hl} and \eqref{eq:hp}).

In Algorithm~\ref{alg:mpnn}, the loop from line~\ref{lin:forT} to line~\ref{lin:mp-end} represents the \emph{message-passing} operations that exchange mutually the information encoded (hidden states)  among links and paths. Likewise, lines \ref{lin:up} and~\ref{lin:mlul} are \emph{update} functions that encode the new collected information into the hidden states respectively for paths and links. The update of paths' states (line \ref{lin:up}) is a simple assignment, while the update of links (line \ref{lin:mlul}) is a trainable neural network. In general, the path update could be also a trainable neural network.

This architecture provides flexibility to represent any source-destination routing scheme. This is achieved by the direct mapping of $\mathcal{R}$ (i.e., the set of end-to-end paths) to specific message passing operations among link and path entities that define the architecture of RouteNet. Thus, each path collects messages from all the links included in it (loop from line~\ref{lin:mp}) and, similarly, each link receives messages from all the paths containing it (line~\ref{lin:mlul}). 
Given that the order of paths traversing the same link does not matter, we used a simple summation for the path-level message aggregation. However, in the case of links, the presence of packet loss may imply sequential dependence in the links that form every path. Consequently, we use a Recurrent Neural Network (RNN) to aggregate link states on paths. Note that RNNs are well suited to capture dependence in sequences of variable size (e.g., text processing). This allows us to model the sequential dependence of links and propagate this information through all the paths.

Moreover, the use of these message aggregation functions (RNN and summation) enables to significantly limit the dimensionality of the problem. The purpose of these functions is to collect an arbitrary number of messages received in every (link or path) entity, and compress this information into fixed-dimension arrays (i.e., hidden states). Note that the size of the hidden states of links and paths are configurable hyper-parameters. In the end, all the hidden states in RouteNet represent an explicit function containing information of the link and path states. This enables to leverage them to infer various features at the same time. Given a set of hidden states $\mathbf h_p^T$ and  $\mathbf h_l^T$, it is possible to connect readout neural networks to estimate some path and/or link-level metrics. This can be typically achieved by using ordinary fully-connected neural networks with some layers and proper activation functions. In Algorithm~\ref{alg:mpnn}, the function $F_p$ (line~\ref{lin:read}) represents a readout function that predicts some path-level features ($\hat{\mathbf y}_p$) using as input the path hidden states $\mathbf h_p$. Similarly, it would be possible to infer some global properties and link-level features ($\hat{\mathbf y}_l$) using also the information in the link hidden states $\mathbf h_l$.

\subsection{Delay, Jitter, and Drops Models}
\label{subsec:delay-jitter-models}

Delay and jitter models are the first published applications of RouteNet showing the capability of this neural architecture to model various network performance metrics. 
In~\cite{routenetsosr2019} these were modeled independently using two neural network models trained to minimize the mean squared error.
Although this approach gives accurate results, it doubles the training time and model parametrization. Also, it hides the fact that average delay and jitter are two statistics of the same random process -- the per-packet delay. In this paper, we propose a generalized probabilistic delay model that can be extended to account for packet drops. 

Formally, the per-path ($i$-th path) delay and jitter are defined as $\mathbb{E} W_i$ and $\mathbb{D}^2 W_i$ respectively.
From the simulation, we obtain sample mean $\bar{w_i}$ and variance $s^2(w_i)$ being their estimates. Instead of modeling $\bar{w_i}$ and $s^2(w_i)$ independently, let us approximate the whole distribution of $W_i$ (marginal distribution given the input features) by a probability distribution parameterized by our RouteNet neural network output $\hat{y}_{i}$ being a two-element vector representing delay and jitter.
Direct generalization of previous models is:

\begin{equation}
W_i \sim Norm(\mu_i, \sigma_i), \quad \mu_i=\hat{y}_{i0}, \enskip \sigma_i = softplus(\hat{y}_{i1}).
\end{equation} 
 Such a model can be trained by maximizing the log-likelihood function of the normal distribution (the loss function is its negative):
 \begin{equation}\label{eq:normloss}
    \ell(\mu_i,\sigma_i)= - n_i\left(\frac{s^2(w_i)}{2\sigma_i^2} + \frac{(\bar{w_i}-\mu_i)^2}{2\sigma_i^2}+\log(\sigma_i)\right),
 \end{equation}
 where $n_i$ is the total number of received packets.
 Note that this loss function is just a scaled squared delay error plus additional terms representing jitter error and it is used in heteroscedastic regression.
Such a simple form of the loss function (negative log likelihood) is possible because the sample mean and variance are the sufficient statistics of the normal distribution and mean-field approximation is used. The per-packet and per-path delays are assumed to be independent and identically distributed (iid) random variables, and the dependence between paths comes from the expected values only. The term mean-field is used because of the similarity of such a model to a mean-field posterior in variational inference. 

In the case of using different distributions to model the delay (e.g., Gamma distribution), we would need to collect different statistics in the training dataset (e.g., $\overline{\log(w_i)}$ for the Gamma distribution) -- namely the sufficient statistics of the distribution. 
In the case of Gamma distribution the loss function would be given by:
\begin{multline*}
	\ell(\alpha_i,\beta_i) = n_i\log\Gamma(\alpha_i) + n_i\beta_i\overline{w_i} +n_i(1-\alpha_i)\overline{\log(w_i)} -\\-n_i\alpha_i\log(\beta_i),
\end{multline*}
where $\alpha$ and $\beta$ would be the RouteNet outputs.

See more details on the derivation of the loss functions for the normal and gamma distributions in Appendix~\ref{appendix:loss}.

Note that this approach is not limited to model only the delay. 
The model can be tuned to different performance characteristics by changing the distribution. Exponential family distributions are perfect tools for this.
In particular, choosing a discrete distribution like Binomial (Poisson is another option) allows us to model per-path packet loss:
\begin{equation}
    L_i\sim Binomial(p_i,n_i+l_i), \quad p_i=sigmoid(\hat{y}_{i}),
\end{equation}
where $p_i$ is the packet loss ratio on path $i$ (i.e., $l_{i}/(n_{i}+l_{i}$).
The log-likelihood function in this case is given by:
\begin{equation}
    \ell(p_i) = l_i\log(p_i)+n_i\log(1-p_i),
\end{equation}
where $l_i$ is the observed number of losses, which is a sufficient statistic for the Binomial distribution. Such a loss function is also common in binary classification problems.

Apart from the introduction of generalized probabilistic modeling to RouteNet, we did not make any other substantial modification with respect to the original implementation in ~\cite{routenetsosr2019}.
The most relevant design choices are:
\begin{enumerate*}[label=\textbf{\arabic*})]
	\item the size of the hidden states for both paths ($\mathbf h_p$) and links ($\mathbf h_l$),
	\item the number of message passing iterations ($T$), and
	\item the neural network architectures for $RNN$, $U$, and $Fp$.
\end{enumerate*}
In line with the previous model, we continue using Gated Recurrent Units (GRU)~\cite{Chung14a}, for both $U$ and $RNN$.
The readout function ($Fp$) is a fully-connected neural network with two layers and uses $selu$ activation functions in order to achieve desirable scaling properties~\cite{Klambauer2017}. 
Compared to the architecture in~\cite{routenetsosr2019}, we added a residual connection from $\mathbf h_p$ to the last hidden layer of the readout function ($Fp$) to provide a direct path for the gradient. This connection shortens the information path from measurement to the model parameters in the message-passing part of the network. This was inspired by the residual connections used in ResNet~\cite{resnet}.

In the readout function, the hidden layers are interleaved with two dropout layers.
The dropout layers play two important roles in the model. During training, they help to avoid overfitting, and during the inference, they can be used for Bayesian posterior approximation~\cite{Gal2015,routenetsosr2019}. 

\section{Baseline}\label{sec:baseline}
To asses the accuracy of RouteNet 
for network performance modeling, we compare it to a queuing theory baseline.
Alternatively to the standard Jackson network model we developed a new approach where each link is modeled as a finite $M/M/1/b$ system instead of $M/M/1$. This allows us to give analytical results for packet loss probability as well as to handle overloaded links. 
In the baseline model we made the following assumptions: 
\begin{enumerate*}[label=\textbf{\arabic*})]
	\item arrival to each queue is approximated by the Poisson process,
	\item packet lengths are approximated by an exponential distribution, and
	\item queues are assumed to be independent.
\end{enumerate*}  
Under those assumptions, we can derive analytical results for queue load, delay distribution, and blocking probability.

Let $\lambda_{k,i}$ be the amount of traffic from path $k$ passing trough link $i$.
For each path we have: 

\begin{equation}\label{eq:qt}
\begin{cases}
\lambda_{k,i}=0\quad \text{if} \quad l_i\not\in p_k\\
\lambda_{k,k(1)}=A_k \\
\lambda_{k,k(j)}=A_k\prod_{i=1}^{j-1}(1-Pb_{k(i)}),\quad j >1 \\
Pb_i=\frac{(1-\rho_i)\rho_i^{b_i}}{1-\rho_i^{b_i+1}}, \\ \rho_i=\frac{\sum_{p_k\in \mathcal{R}}\lambda_{k,i}}{c_i}
\end{cases}
\end{equation}
where 
$\sum_{p_k\in \mathcal{R}}\lambda_{k,i}$ is the total traffic on the $i$-th link, $Pb$ denotes the blocking probability, $A_i$ is the demand on the $i$-th path, $b$ is the buffer size, and $c$ is the link capacity.
The system of equations~\eqref{eq:qt} is derived from traffic balance on the lossy network and it is solved using the fixed-point method.
In the first iteration, we assume no packet loss to compute initial traffic intensities.
Given the first approximation, we can compute the loss probability and update the intensities to account for the losses. 
After a few iterations, the algorithm converges to a fixed point. 
After convergence, it is possible to compute the delay, jitter and drop probability for each link from standard queuing theory~\cite{Kelly2011}.
Path statistics are computed assuming link independence. 
Notice how this approach is similar to our derivation of the neural architecture of RouteNet.
In the baseline, we use known relations from queuing theory while in RouteNet those relations are approximated by a neural network and learned from the data. 

\section{Evaluation of the Accuracy of the GNN Model}

In this section, we evaluate the accuracy of RouteNet (Sec. \ref{sec:modeling-gnn}) to estimate the per-source/destination mean delay/jitter and the number of packet drops in a wide variety of network topologies, routing schemes and traffic intensities.

\subsection{Simulation Setup}
\label{subsec:simulation-setup}

We built a ground truth for our GNN model with a custom-built packet-level simulator with queues using OMNeT++ v4.6~\cite{omnet}. Each simulation, we compute the mean end-to-end delay and jitter, and the packets dropped for every source-destination pair along 16k time units. We model the traffic exchanged by every src-dst pair with the following traffic matrix ($\mathcal{TM}$):
\begin{equation}
\mathcal{TM}(S_i,D_j)~=~\frac{\mathcal{U}(0.1, 1)*TI}{N-1}\quad \forall~i,j \in nodes, i\neq j\label{eq:traffic-matrix}
\end{equation}

Where $\mathcal{U}(0.1, 1)$ is a uniform distribution in the range \mbox{[0.1, 1]}, \textit{TI} represents a tunable parameter of the overall traffic intensity in the simulation, and \textit{N} is the number of nodes in the network topology. In each source-destination pair, inter-packet arrival times are modeled with an exponential distribution whose mean is derived from the traffic defined in $\mathcal{TM}$. Also, packet sizes follow a binomial distribution, where 50\% of the packets have a size of 300 bits and the rest of packets contain 1700 bits. All the queues have a size of 32 packets. We made simulations in 4 different topologies with variable link capacity and traffic intensity. Link capacities range the following values: 10, 40 or 100 kbps.

\vspace{-0.2cm}
\subsection{Training and Evaluation}
\label{subsec:traning-and-evaluation}

We implemented both models (delay and loss) in TensorFlow.
The source code and all the training/evaluation datasets used in this paper are publicly available at~\cite{kdngit}. The current implementation is heavily optimized in terms of performance. The training speed was improved by a factor of 10x compared to the first version reported in~\cite{routenetsosr2019}.
This allowed us to train the model on a considerably larger dataset consisting of samples from the 14-node NSF~\cite{nsfnet}, 24-node Geant2~\cite{geant2} and 50-node Germany50~\cite{SNDlib10} networks.

In total all models (delay/jitter, and drops) were trained on a collection of 260,000  samples.
Despite this dataset contains only samples from the three topologies mentioned above, it includes over 200 different routing schemes and a wide variety of traffic matrices with different traffic intensity. For testing (during training), we use 112,000 samples.

In our experiments, we select a size of 32 for both the path's hidden states ($\mathbf h_p$) and the link's hidden states ($\mathbf h_l$). The initial path features ($\mathbf x_p$) are defined by the bandwidth that each source-destination path carries (extracted from the traffic matrix $\mathcal{TM}$) while the initial link features ($\mathbf x_l$) are the link capacities. 
Note that, for larger networks, it might be necessary to use larger sizes for the hidden states. Moreover, every forward propagation we execute $T$=8 iterations. 
The dropout rate is equal to 0.5. This means that each training step we randomly deactivate half of neurons in the readout neural network. 
This also allows us to make a probabilistic sampling of results and infer the confidence of the estimates.

\begin{figure*}[!t]
 \centering
 \subfloat[RouteNet\label{subfig:eval_all}]{%
   \includegraphics{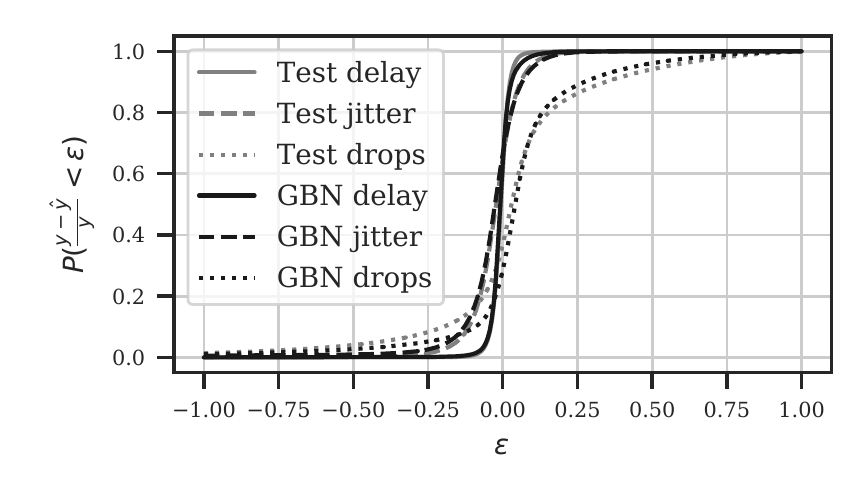}
   \vspace{-0.2cm}
 }
 \subfloat[Baseline\label{subfig:baseline}]{%
   \includegraphics{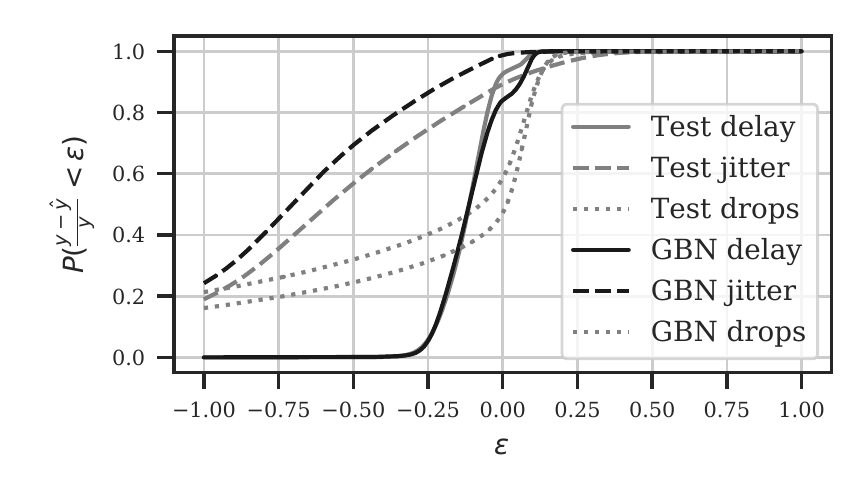}
   \vspace{-0.2cm}
 }
 
 \caption{Cumulative Distribution Function (CDF) of the relative error. Solid line for delay, dashed line for jitter, dotted line for packet loss ratio (only in the cases with observed drops). $y$ is the true value, while $\hat y$ denotes the model prediction.}
 \label{fig:eval_all}
 \vspace{-0.3cm}
\end{figure*}

During the training we minimized the loss function of each model (negative log likelihood of the target distribution) between the predictions of RouteNet and the ground truth plus the $L2$ regularization loss (weight decay 0.1). 
The loss was summed over all source/destination pairs. 
We introduce a minibatch of samples by using a single disconnected graph composed by the individual connected graphs in the batch.
The total loss function is minimized using an Adam optimizer with an initial learning rate of 0.001.

We executed the training over 260,000 batches of 16 samples randomly selected from the training set. 
In our testbed with a GPU Nvidia  GeForce GTX 1080, this took around 20 hours ($\approx$70 samples per second)\footnote{Note that this time consuming operation is required only once.
	The inference is many orders of magnitude faster. On the same hardware the single inference for a network with 200 nodes and 39,800 paths takes $99.2\text{ ms}\pm 561\text{ }\mu s $, while on CPU (i5-6400) it takes  $1.26\text{ s}\pm 9.57\text{ }ms $.}.

\begin{table}[!t]
	\caption{Summary of the evaluation results (MRE); RN -- RouteNet, QT --  Queuing Theory}
	
	\label{tab:eval}
	\centering
\begin{tabular}{l|ll|ll|ll}
            \toprule
           & \multicolumn{2}{l|}{\bfseries Delay} & \multicolumn{2}{l|}{\bfseries Jitter} & \multicolumn{2}{l}{ \bfseries Drops} \\\midrule
           & \bfseries RN       & \bfseries QT      & \bfseries RN       & \bfseries QT      & \bfseries RN       & \bfseries QT       \\\midrule
\bfseries Test set & 0.022             &   0.124          &  0.061            &  0.615           &   0.180           &    0.713       \\
\bfseries GBN*        & 0.025        & 0.126      & 0.078        & 0.720       & 0.154        & 0.556    
 
\end{tabular}
\vspace{-0.4cm}
\end{table}

Table~\ref{tab:eval} shows a summary of the delay, jitter, and loss experiments we made in 4 different network topologies. 
We report the Mean Relative Error (MRE) for both the RouteNet (RN) and the queuing theory baseline (QT).
For the NSF, Geant2 and Germany50 networks, we evaluate the accuracy over the same 112,000 samples used for testing during the training process (test set). Additionally, we tested the accuracy over a dataset 
simulated in the 17-node GBN topology~\cite{gbn} with 87,000 samples. 
Note that \emph{the GBN network was never included in the training.}
The model was only trained with samples from the \emph{NSF (14 nodes), Geant2 (24 nodes), and Germany50 (50 nodes)} networks.
The high accuracy on the \emph{GBN} network (17 nodes) reveals the ability of RouteNet to well generalize even to new networks.
In all cases, RouteNet outperforms the queuing theory baseline, what is rather remarkable given the Poisson traffic distribution used in the simulator. For this traffic, the queuing theory should be a quite accurate approximation, yet the machine learning model achieves better accuracy. 

Statistics like MRE provide a good picture of the general accuracy of the model. 
However, there are more elaborated methods that offer a more detailed description of the model behavior. 
Hence, we focus on the full distribution of residuals (i.e., the error of the model and the baseline). 
Particularly, we present a CDF of the relative error (Fig.~\ref{fig:eval_all}) over all the evaluation samples. 
This allows us to provide a comprehensive view of the whole evaluation in a single plot. 
In these results, we can observe that the prediction error in general is considerably low. 
Moreover, we see that the drops model is more biased compared to the delay model. 
Note that the plot contains only the cases where we observed one or more packets dropped, otherwise we would face division by zero when computing the relative error.
However, in our simulation datasets there are many cases with zero drops, and these cases are also accurately predicted by the RouteNet model. To test this one can compute the correlation coefficient between predictions and true loss ratio (including cases without loss where the value is 0).
In all experiments it is $\geq0.997$ for both the test and evaluation sets, and also for the baseline. This number is high because cases with zero drops were always correctly labeled with a small probability.

The generalization to unknown routings and traffic matrices in known topologies (NSF, Geant2 and Germany50) is almost perfect.
The model is also equally accurate for the unknown topology (GBN). 
This reveals the possibility to deploy RouteNet models in different network scenarios where they were not trained. Also, there is the possibility to fine-tune only the readout part with samples of the new networks and reuse the computationally intensive message-passing part.

\subsection{Generalization Capabilities}
This section discusses the generalization capabilities and limitations of RouteNet. As in all ML-based solutions, RouteNet is expected to provide more accurate inference as the distribution of the input data is closer to the distribution of training samples. In our case, it involves topologies with similar number of nodes and distribution of connectivity, routing schemes with similar patterns (e.g., variations of shortest path), and similar ranges of traffic intensities. We experimentally observe the capability of RouteNet to generalize to topologies of variable size (from 14 to 50 nodes) while still providing accurate estimates. In order to expand the generalization capabilities of RouteNet, an extended training set must be used including a wider range of distributions of the input elements.

RouteNet's architecture is built to estimate path-level metrics using information from the output path-level hidden states. However, it is relatively easy to modify the architecture and use information encoded in the link hidden states to produce link-related metrics inference (e.g., congestion probability on links).

\section{Use Cases}

This section shows two different use cases where we leverage the predictions of RouteNet (Sec. \ref{sec:modeling-gnn}) to address relevant network optimization tasks from the control plane. In these use cases we use the delay, jitter and drops models of RouteNet to evaluate the resulting performance after applying some modifications in the network configuration. Particularly, we limit the optimization problem to : \mbox{$(i)$ generate} a set of candidate configurations (e.g., routing schemes), \mbox{$(ii)$ evaluate} the resulting performance for each of them, and \mbox{$(iii)$ select} the one that best fits the optimization objective. We compare the performance achieved by our optimizer based on RouteNet to the results obtained by classic optimizers based on link utilization, the widely deployed Shortest Path routing policy, and the optimal solution using an accurate packet-level simulator.

In this context, state-of-the-art models predicting Key Performance Indicators (KPI) such as delay, jitter or drops are not suited to perform online network optimization at large scale, since they often result into inaccurate estimation (e.g., analytic models) and/or prohibitive processing cost (e.g., packet-level simulators). All the evaluations in this section are performed in network scenarios of the NSF network topology~\cite{nsfnet}. For the RouteNet-based optimizer we use the delay/jitter and drops models trained on the NSF, Geant2 and Germany50 datasets (Sec.~\ref{subsec:traning-and-evaluation}).

\subsection{Delay, Jitter, and Loss-aware Routing Optimization}

\begin{figure*}[!t]
 \centering
 \subfloat[Mean delay\label{subfig:mean-delay}]{%
   \includegraphics[width=0.3\linewidth]{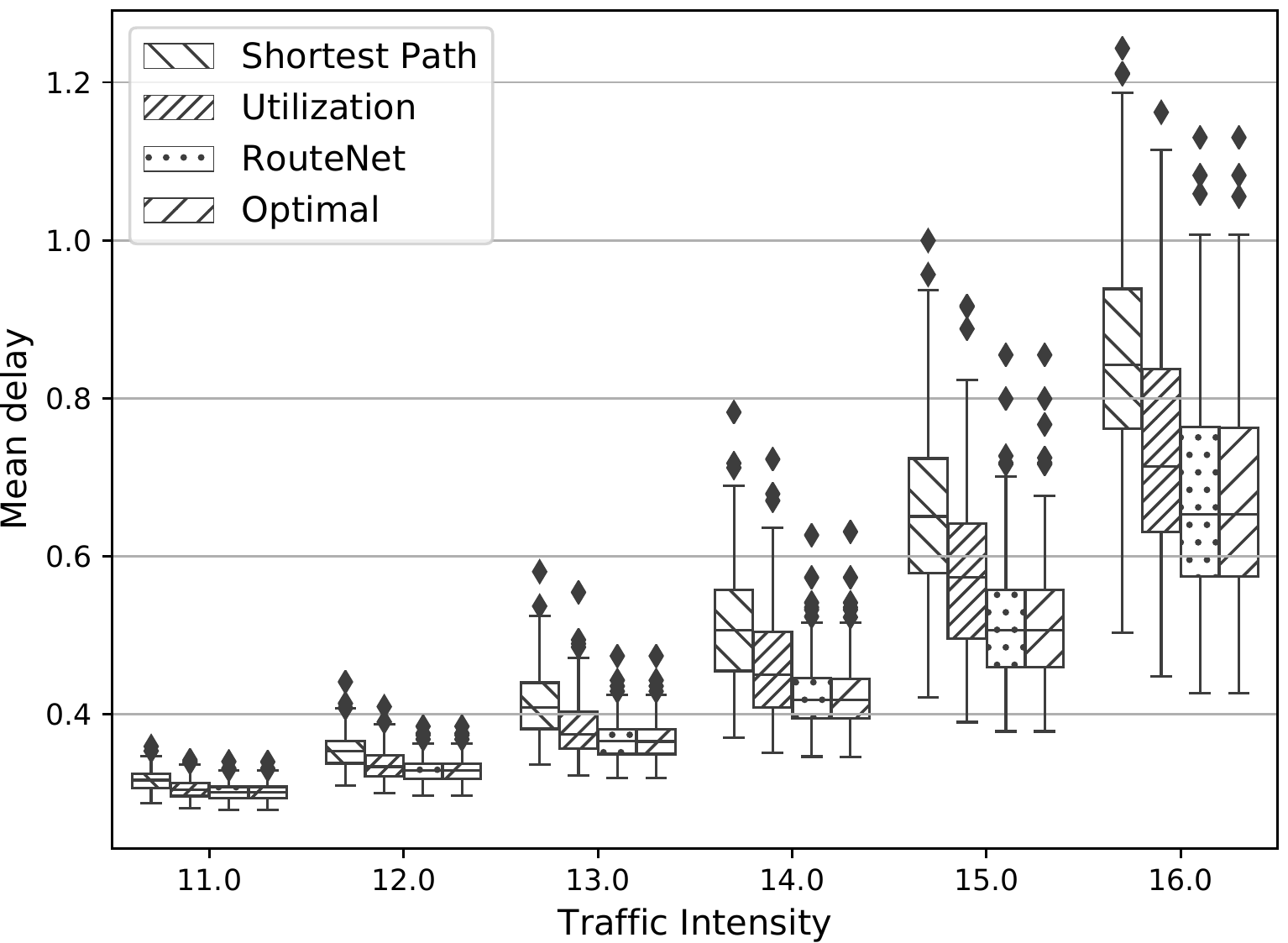}
 }
 \hfill
 \subfloat[Packet loss\label{subfig:packet-loss}]{%
   \includegraphics[width=0.3\linewidth]{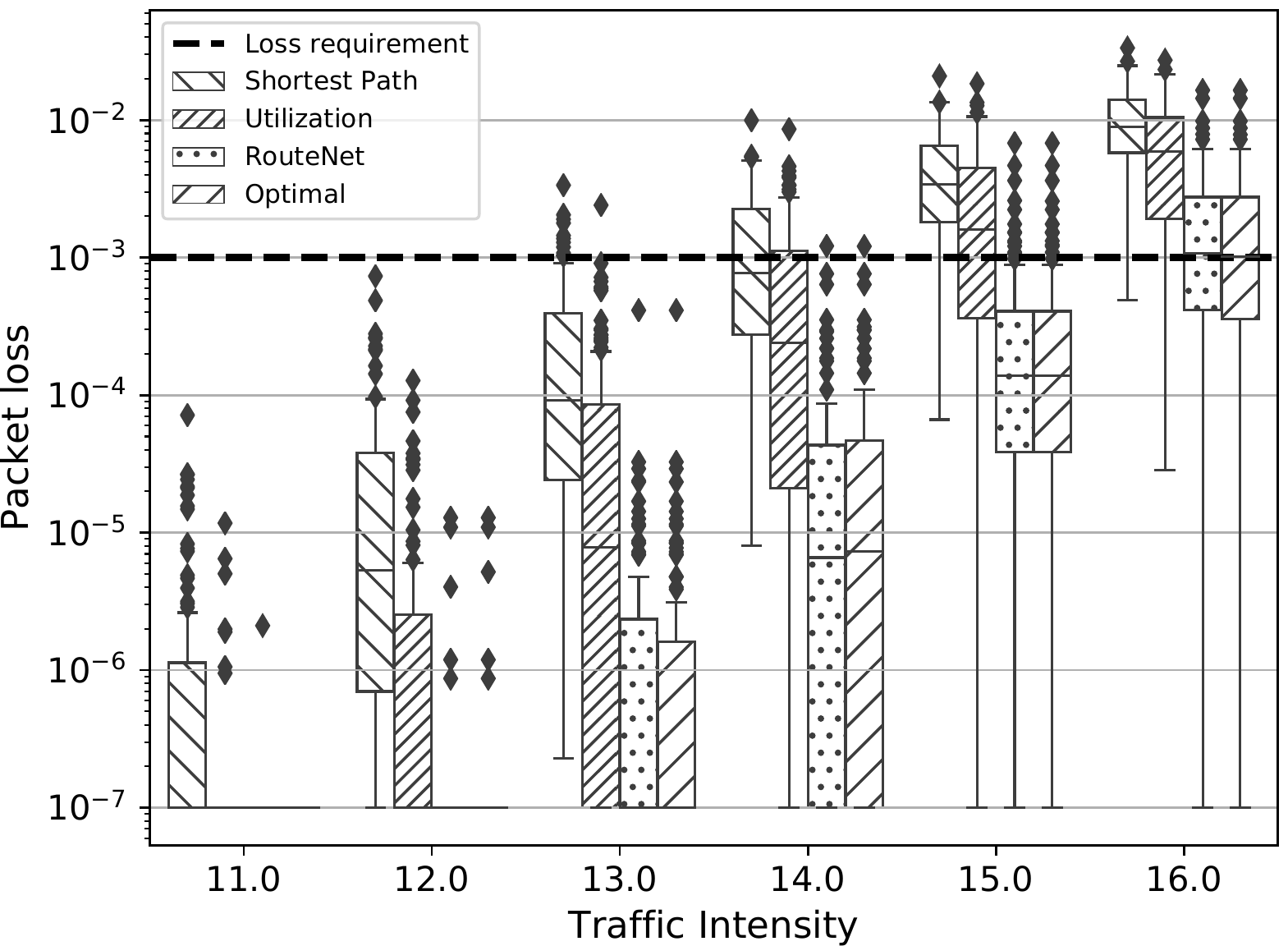}
 }
 \hfill
 \subfloat[Mean jitter/mean delay\label{subfig:mean-jitter}]{%
   \includegraphics[width=0.3\linewidth]{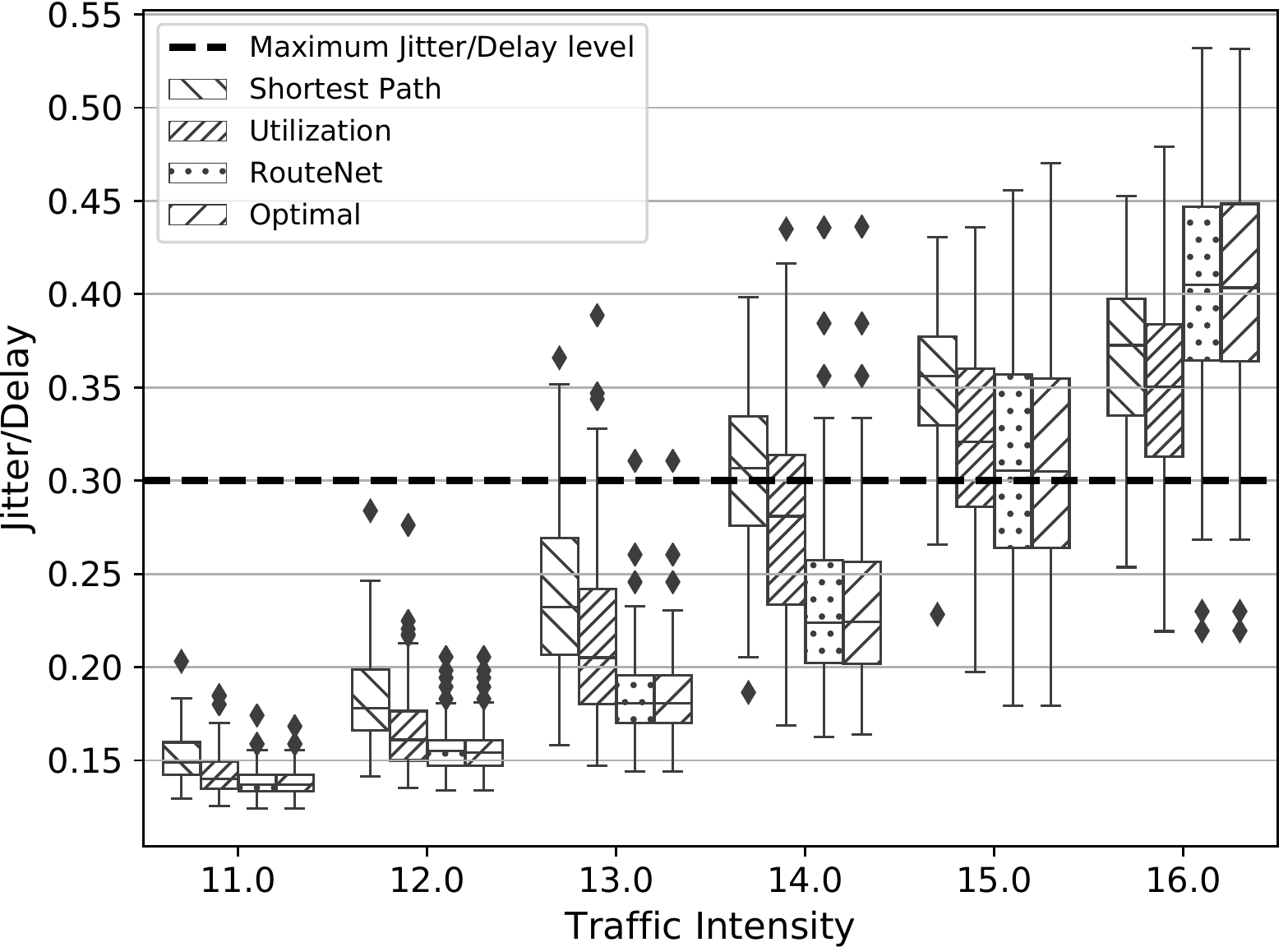}
 }
 \caption{Evaluation of the delay,jitter, and loss-aware routing optimization use case.}
 \label{fig:evaluation-routing-opt}
\end{figure*}

This use case represents a QoS-aware routing optimization scenario where the target policy is to make a joint optimization of multiple KPI. Particularly, we leverage the KPI predictions of RouteNet to minimize the per-source/destination mean delay and guarantee at the same time that jitter and packet loss are below certain thresholds. We define the following optimization objectives in decreasing order of priority:
\begin{enumerate}
    \item Maintain the mean packet loss rate below 0.1\%.\\
    $\left[i.e., mean(L_{i}/n_{i})<10^{-3}\right]$
    \label{condition-1}
    
    \item Maintain the mean per-source/destination jitter below 20\% of the mean delay\\$\left[i.e., mean(jitter/delay)<0.2\right]$
    \label{condition-2}
    
    \item Minimize the mean per-source/destination delay in the network
    \label{condition-3}
\end{enumerate}

We implemented a RouteNet-based optimizer that utilizes the performance predictions made by RouteNet (i.e., mean delay, jitter, and loss). After evaluating a given set of candidate routing schemes, this optimizer selects the one that better fulfills the optimization objectives according to the RouteNet predictions. In particular, it selects the routing scheme that results in lower per-source/destination mean delay among those configurations that fulfill the loss and jitter constraints (\ref{condition-1} and \ref{condition-2}). In the case that no routing scheme satisfies the packet loss restriction (\ref{condition-1}), the optimizer selects the routing that minimizes the mean packet loss regardless of the other metrics. Likewise, if there is no routing that satisfies the jitter restriction (\ref{condition-2}), then the optimizer selects the routing scheme with lower mean delay among those that still satisfy the loss constraint (\ref{condition-1}). The set of candidate routing configurations comprises 450 variants of the shortest path policy. To this end, we consider an initial scenario where all the links of the NSF topology have a weight equal to 1 and we run the Dijkstra algorithm to compute the shortest path configuration. Then, the remaining 449 routing variations are generated by adding 0.05 to the weight of 21 links randomly selected. Note that in this random selection process a link can be chosen more than once.

We compare the results obtained by our RouteNet-based optimizer with two traditional routing approaches: $(i)$ Shortest Path routing (hereafter SP), and $(ii)$ a more elaborated routing optimizer based on link utilization. For the SP routing baseline, we compute the resulting performance after applying the 450 routing candidate solutions, which \mbox{ -- as} mentioned \mbox{before --} are all variants of the shortest path. Thus, the evaluation results for this baseline represent the average performance achieved over all the SP routing variants (i.e., avg. delay, jitter and loss over the 450 configurations). The utilization-based optimizer selects the routing configuration that results in a more balanced link utilization (i.e., less variance over all the links' utilization). Similarly to state-of-the-art \mbox{utilization-based} routing strategies, in this case we consider a fluid model of the network without considering loss to compute the resulting link utilization. Note that alternative utilization-based criteria could also be applied, such as minimizing the utilization of the most loaded link. Moreover, we compute the optimal solution with an optimizer that relies on the accurate performance estimates produced by our packet-level simulator (Sec.~\ref{subsec:simulation-setup}). In other words, this latter optimizer evaluates all the possible routing configurations with our packet-level simulator and selects the one that best fits the optimization targets. For a fair comparison, all the optimizers consider the same set with 450 routing schemes.

We evaluate the performance achieved by all the routing strategies in scenarios with variable traffic intensity (from low to high load). Particularly, we consider 6 different traffic intensity levels. 

Figure~\ref{fig:evaluation-routing-opt} summarizes the average per-source/destination mean delay (Fig.~\ref{subfig:mean-delay}), average packet loss (Fig.~\ref{subfig:packet-loss}), and average jitter/delay ratios (Fig.~\ref{subfig:mean-jitter}) obtained by the different optimizers with respect to the traffic intensity (x-axis). Note that each boxplot represents the results over 100 scenarios with different input traffic matrices of the same traffic intensity (TI). To this end, we generated 100 traffic matrices ($\mathcal{TM}$) for each TI (from 11 to 16) according to Equation~\eqref{eq:traffic-matrix}. For all the optimization strategies, we provide the resulting performance metrics computed by our packet-level simulator after applying the best routing configuration selected in each case. To display packet loss (Fig.~\ref{subfig:packet-loss}), we use a logarithmic scale (y-axis). Since there are some cases without any loss (mostly at lower traffic intensities), we defined a lower limit to avoid minus infinity values ($packet\_loss=10^{-7}~\forall~packet\_loss\leq 10^{-7}$). 

Looking into Figures~\ref{subfig:packet-loss} and~\ref{subfig:mean-jitter}, the most remarkable result is that the optimizer based on RouteNet's predictions was able to maintain the loss and jitter constraints even in most of the scenarios with medium-high traffic load (TI=14). Only in some cases with the highest traffic intensity \mbox{(TI=15-16)} it did not find any routing scheme meeting the loss requirement. In contrast, the traditional SP policy and the utilization-based optimizer start to exceed the loss threshold from TI=13 (medium load). Moreover, in Figure~\ref{subfig:mean-delay} we observe that the RouteNet-based optimizer clearly outperforms these other optimizers also in terms of per-source/destination mean delay. Particularly, as the network scenarios become more challenging (i.e., higher traffic intensity) the difference in performance is more remarkable. Note that in the cases where the loss requirement could not be fulfilled, the RouteNet-based optimizer selected the routing scheme that resulted in less packet loss regardless of the mean delay and jitter predictions. However, in these cases the configuration with lower loss resulted also in lower average delay and jitter compared to traditional routing techniques.

Additionally, we evaluated the performance achieved by the optimizer that uses directly the delay, jitter, and loss metrics computed by our packet-level network simulator (Sec.~\ref{subsec:simulation-setup}). These results are labeled as ``optimal'' in Figures~\ref{subfig:mean-delay},~\ref{subfig:packet-loss},~and~\ref{subfig:mean-jitter}. As we can observe, the resulting performance using the accurate predictions of RouteNet is practically the same as when we use the network simulator metrics. This illustrates the potential of RouteNet to be used for network optimization offering similar performance than computationally intensive optimizers based on packet-level simulation.

\subsection{Budget-constrained Network Upgrade}

\begin{table*}[!t]
\caption{Evaluation results of the Optimal link placement use case}
\resizebox{\textwidth}{!}{%
\begin{tabular}{c|cc|ccc|ccc|cc}
\toprule
    \multirow{2}{*}{{\textbf{\begin{tabular}[c]{@{}c@{}}Traffic\\matrix\end{tabular}}}}&\multicolumn{2}{c|}{\textbf{Original scenario}} & \multicolumn{3}{c|}{\textbf{RouteNet-based optimizer}} & \multicolumn{3}{c|}{\textbf{Baseline}} & \multicolumn{2}{c}{{\textbf{\begin{tabular}[c]{@{}c@{}}Relative reduction\\RouteNet-based vs Baseline\end{tabular}}}}\\ \cline{2-11}
     & \textbf{Delay} & \textbf{Jitter/delay} & \textbf{Opt. link placement} & \textbf{New delay} & \textbf{New jitter/delay} & \textbf{Most loaded link} &\textbf{New delay} & \textbf{New jitter/delay} & \textbf{Rel. Delay (\%)} & \textbf{Rel. jitter/delay (\%)}\\ \midrule \midrule
     $TM_{1}$ & 0.964 & 0.345 & (11-1) & 0.715 & 0.222 & (3-8) & 1.377 & 0.352 & 48.09\% & 36.77\% \\
     $TM_{2}$ & 0.909 & 0.335 & (9-2) & 0.498 & 0.223 & (12-5) & 1.065 & 0.310 & 53.22\% & 28.12\% \\
     $TM_{3}$ & 0.949	& 0.374 & (11-2) & 0.752 & 0.236 & (3-8) & 1.312 & 0.320 & 42.66\% & 26.10\% \\
     $TM_{4}$ & 1.130 & 0.339 & (12-2) & 0.684 & 0.205 & (5-12) & 1.387 & 0.293 & 50.70\% & 29.95\% \\
     $TM_{5}$ & 0.967 & 0.314 & (10-0) & 0.630 & 0.214 & (5-12) & 1.321 & 0.311 & 52.33\% & 31.08\% \\
     $TM_{6}$ & 1.007 & 0.362 & (11-1) & 0.764 & 0.206 & (12-5) & 1.192 & 0.298 & 35.90\% & 30.97\% \\
     $TM_{7}$ & 1.055 & 0.379 & (9-2) & 0.743 & 0.223 & (12-5) & 1.087 & 0.345 & 31.67\% & 35.41\% \\
     $TM_{8}$ & 0.955 & 0.360 & (11-1) & 0.749 & 0.208 & (3-8) & 1.281 & 0.310 & 41.52\% & 32.90\% \\
\bottomrule
\end{tabular}
}
\label{tab:newLink}
\end{table*}

This use case addresses a well-known optimization problem in networking: how to optimally upgrade the network by adding new links in the topology.

For this use case, we selected 8 different network scenarios from the previous use case where the RouteNet-based optimizer could not meet the loss requirement (0.1\%) given the high traffic load. In particular, these scenarios include traffic matrices (TM) of the highest load (TI=16). For each scenario, the RouteNet-based optimizer selects the optimal link placement in combination with the routing scheme that results in lower per-source/destination mean delay. Particularly, we limit the problem to add only one link of 10 kbps, which is the minimum link capacity considered in the NSF topology. Then, the optimizer evaluates all the possible link placements in the NSF network. Moreover, we consider 450 different routing schemes for each new possible link placement. The routing configurations are generated using the same method as in the previous use case, i.e., they are shortest path variants. We compare these results with the use of a traditional strategy that network operators often apply when they detect degradation in network performance. Particularly, this strategy selects the most loaded link (i.e., with highest utilization) in the current scenario and replaces it with another link with more capacity. In our evaluation setup, the NSF network originally contains links of 10 and 40 kbps. Then, if the most loaded link has a capacity of 10 kbps, it is replaced by a 40 kbps link. Likewise, 40 kbps links are substituted by 100 kbps links.

Table~\ref{tab:newLink} shows the optimal link placement in the NSF network topology under the 8 TMs of high traffic intensity (TI=16). For each TM, we show the average delay and the jitter/delay ratio before and after adding the optimal link with the best routing configuration. Also, we include the results applying the approach that updates the link capacity of the most loaded link (labeled as ``Baseline''). All these results show the performance metrics computed by our packet-level simulator. To this end, we simulate the new scenarios considering the optimal link placement and routing scheme selected by the RouteNet-based optimizer and the link upgrade selected by the baseline. Here, we can observe that the optimizer using RouteNet achieves an important reduction on the mean delay. In particular, it achieves on average $\approx$44.5\% more reduction in delay than the baseline. Note that the optimization target is only based on minimizing the mean delay. However, we also provide the results for jitter. As we expected, there is also an important reduction on this metric. We can observe that the jitter/delay ratio is reduced $\approx$31.4\% on average with respect to the baseline.

\section{Related Work}
The ultimate goal of network modeling can be summarized as to provide a cost function for optimization. Many attempts have been done over the years to derive the perfect solution. Those include discrete-state Markov models (queuing theory), stochastic fluid models and network calculus. Among them, queuing theory is the most popular. 
The most advanced approach to network optimization constructs an objective function based on the linearization of well known queuing theory results~\cite{Pioro2004}.
Network calculus is used for the worst-case scenario in networks, so it cannot be directly compared to RouteNet as those worst cases are rarely observed in operational environments. Although fluid models are efficient and popular for congestion control they are by design approximate and may lead to inaccurate results due to hiding the inherent properties of the traffic~\cite{Eun2007}.

Given that deep learning models can learn queuing theory with high accuracy~\cite{rusek2018message}, we may want to leverage these models for network optimization tasks. The main advantage of this kind of models is that they always benefit from new data. Every edge case can be used to improve the model with minimal investment.

Network modeling with deep neural networks is a recent topic proposed in the literature \cite{wangMachineLearning,kdn} with few pioneering attempts. The closest works to our contribution are first \mbox{Deep-Q} \cite{deepQ}, where the authors infer the QoS of a network using the traffic matrix as an input using Deep Generative Models. And second \cite{mestresModeling}, where a fully-connected feed-forward neural network is used to model the mean delay of a set of networks using as input the traffic matrix. The main goal of the authors is to understand how fundamental network characteristics (such as traffic intensity) relate with basic neural network parameters (depth of the neural network). RouteNet is also able to produce accurate estimates of performance metrics (delay, jitter and loss), but it does not assume a fixed topology and/or routing, rather it is able to produce such estimates with arbitrary topologies and routing schemes not seen during training. This enables RouteNet to be used for network operation, optimization, and what-if analysis. 

Finally, an early attempt to use Graph Neural Networks for computer networks can be found in \cite{geyer2018learning}. In this case the authors use a GNN to learn shortest-path routing and \mbox{max-min} routing using supervised learning. While this approach is able to generalize to different topologies it cannot generalize to different routing schemes beyond the ones for which it has been specifically trained. In addition, the focus of the paper is not to estimate the performance of such routing schemes. Another paper of the same author is focused on performance~\cite{Geyer2019}, however, the model is a standard Graph Neural Network, which loses information about the order of queues on paths. In contrast, RouteNet is designed to explicitly use this information. The importance of order can be found in the derivation of the baseline in section~\ref{sec:baseline}.

\section{Conclusions}

Software-Defined Networks offer an unprecedented degree of flexibility in network control and management that, combined with timely network measurements collected from the data plane, open the possibility to achieve efficient online network optimization.

However, existing network modeling techniques based on analytic models (e.g., Queuing theory) cannot handle this huge complexity. As a result, current optimization approaches are limited to improve a global performance metric, such as network utilization or planning the network based on the worst case estimates of the latencies obtained from network calculus.

In this context, Deep Learning is a promising solution to handle such complexity and to exploit the full potential of the SDN paradigm. However, earlier attempts to apply Deep Learning to networking problems resulted in tailor-made solutions that failed to generalize to other network scenarios.

In this paper, we presented RouteNet, a custom architecture based on Graph Neural Network (GNN) specifically designed for computer network modeling. RouteNet uses a novel message-passing function that allows the GNN to capture the complex relationships between the state of paths and links resulting from network topologies and routing configurations in order to model the resulting network performance.

We designed and implemented an extended RouteNet model based on Generalized Linear Models that predicts the distribution of the per-source/destination per-packet delay and loss in networks. From these output distributions we evaluate the accuracy of the mean per-packet delay, the jitter, and the mean packet loss predicted. Our evaluation results show that RouteNet is able to generalize to other network topologies, routing configurations and traffic matrices not seen in the training.

Also, the modular architecture of RouteNet simplifies transfer learning, which consists of reusing neural network models trained for a particular task and retrain them to address other problems in similar domains. In the context of RouteNet this was proposed in~\cite{routenetsosr2019}, where the jitter model was bootstrapped from an early stage of the delay model.

Lastly, we presented some optimization use cases where we use the performance predictions of RouteNet for different network optimization purposes. In particular, we perform \mbox{QoS-aware} routing optimization based on delay, jitter and packet loss requirements, and also use RouteNet to find the optimal link placement in a network planning scenario.

\bibliographystyle{IEEEtran} 
\bibliography{JSAC_manuscript}
\enlargethispage{-3.5in}
\begin{IEEEbiography}[{\includegraphics[width=1in,height=1.25in,clip,keepaspectratio]{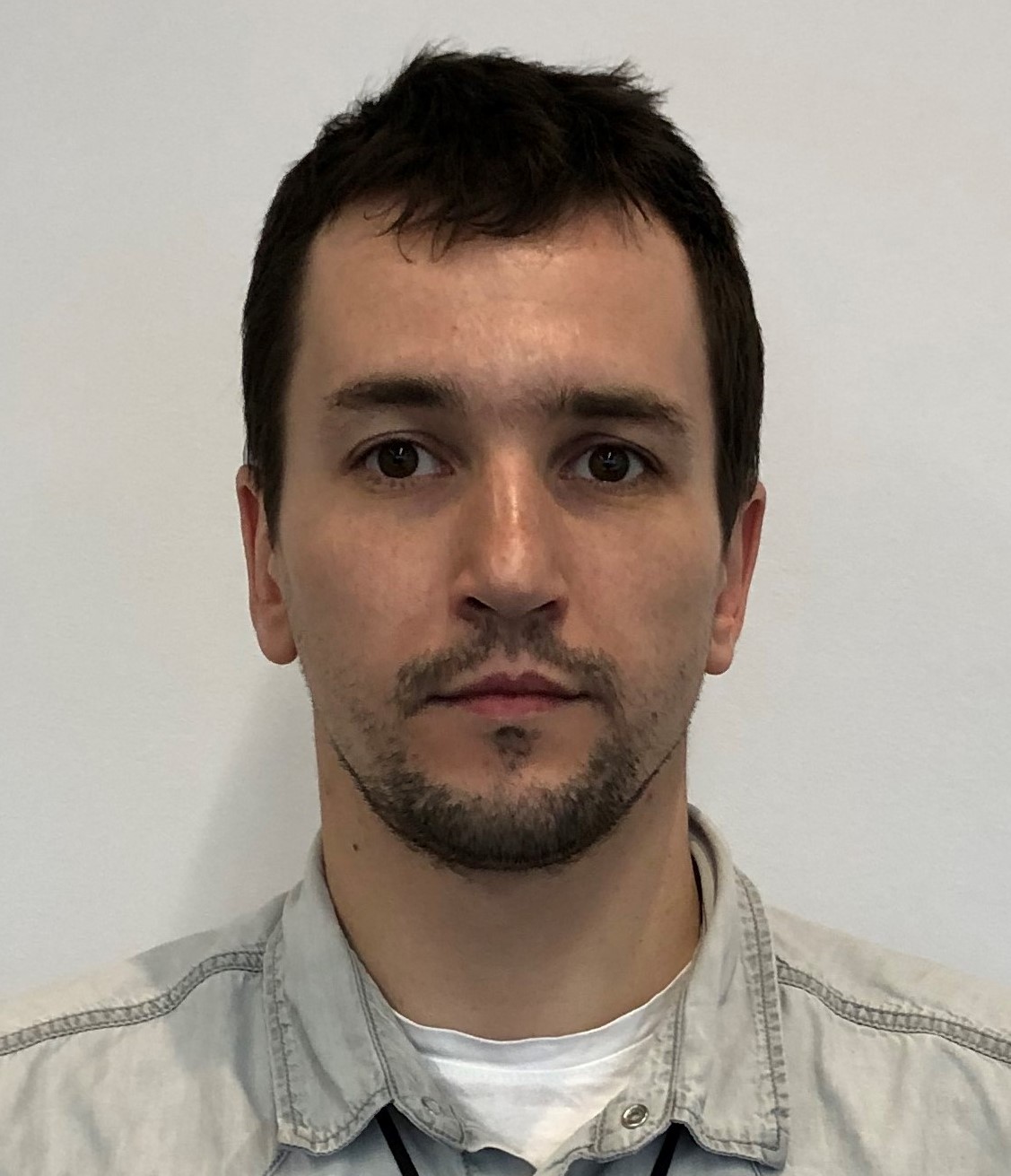}}]%
{Krzysztof Rusek}
is an assistant professor at  AGH and data scientist at the Barcelona Neural Networking Center.
He defended his Ph.D. Thesis on queuing theory in 2016 at AGH.  
Prior to that he has worked as a system administrator and machine learning engineer in the research group focused on processing and protection of multimedia content.  
His main research interests are performance evaluation of telecommunications systems, machine learning and data mining. 
Currently, he is working on the applications of Graph Neural Networks and probabilistic modeling for performance evaluation of communications systems and data mining in Astronomy.
\end{IEEEbiography}
\begin{IEEEbiography}[{\includegraphics[width=1in,height=1.25in,clip,keepaspectratio]{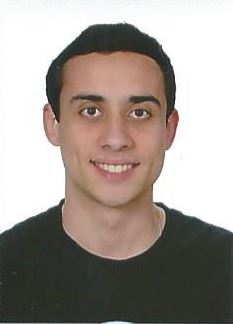}}]%
{Jos\'e~Su\'arez-Varela}
received his B.Sc. and M.Sc. degrees in Telecommunication engineering from the Universidad de Granada (UGR), in 2014 and 2017 respectively. He is currently a Ph.D. candidate at the Barcelona Neural Networking Center (BNN-UPC). During 2019, he was a visiting researcher at the University of Siena. His main research interests are in the field of Artificial Intelligence applied to networking, particularly on the application of Graph Neural Networks for network modeling and optimization. He is also interested in traffic measurement and classification, and their application in Software-Defined Networking.
\end{IEEEbiography}
\newpage
\enlargethispage{-3in}
\begin{IEEEbiography}[{\includegraphics[width=1in,height=1.25in,clip,keepaspectratio]{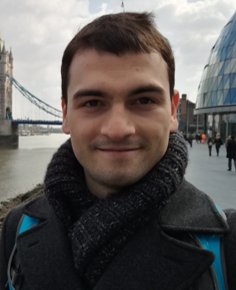}}]%
{Paul Almasan}
received his B.Sc. and M.Sc. in Computer Science from the Universitat Polit\`ecnica de Catalunya (UPC), Spain, in 2017 and 2019 respectively. He is currently pursuing his Ph.D. degree at the Barcelona Neural Networking Center (BNN-UPC). His research interests are focused on Graph Neural Networks and Deep Reinforcement Learning applied to network optimization.
\end{IEEEbiography}
\begin{IEEEbiography}[{\includegraphics[width=1in,height=1.25in,clip,keepaspectratio]{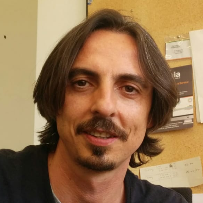}}]%
{Pere Barlet-Ros}
is an associate professor at Universitat Polit\`ecnica de Catalunya (UPC) and scientific director at the Barcelona Neural Networking Center (BNN-UPC). From 2013 to 2018, he was co-founder and chairman of the machine learning startup Talaia Networks. The company was acquired by Auvik Networks in 2018. He was also a visiting researcher at Endace (New Zealand), Intel Research Cambridge (UK) and Intel Labs Berkeley (USA). His research interests are in machine learning technologies for network management and optimization, traffic classification and network security. In 2014, he received the 2nd VALORTEC prize for the best business plan awarded by the Catalan Government (ACCIO) and in 2015 the Fiber Entrepreneurs award as the best entrepreneur of the Barcelona School of Informatics (FIB).
\end{IEEEbiography}

\begin{IEEEbiography}[{\includegraphics[width=1in,height=1.25in,clip,keepaspectratio]{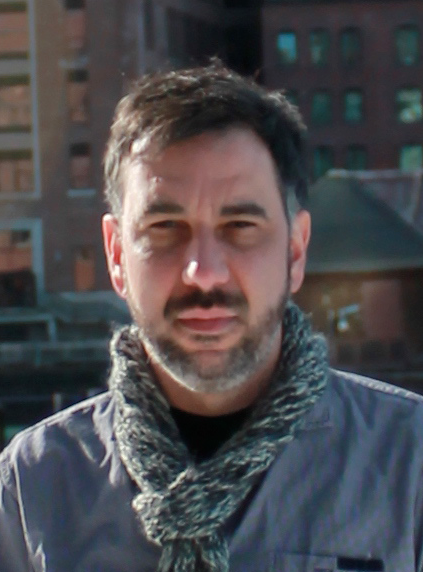}}]%
{Albert Cabellos-Aparicio}
is an assistant professor at Universitat Polit\`ecnica de Catalunya (UPC), where he obtained his PhD in computer science engineering in 2008. He is director of the Barcelona Neural Networking Center (BNN-UPC) and scientific director of the NaNoNetworking Center in Catalunya. He has been a visiting researcher at Cisco Systems and Agilent Technologies, and a visiting professor at the KTH, Sweden, and the MIT, USA. His research interests include the application of Machine Learning to networking and nanocommunications. His research achievements have been awarded by the Catalan Government, his university, and INTEL. He also participates regularly in standardization bodies such as the IETF.
\end{IEEEbiography}

\newpage
\appendix[Loss functions]
\label{appendix:loss}

\subsection{Normal}\label{normal}
The pdf of the normal distributions is:
\[f(x) = \frac{1}{\sigma \sqrt{2\pi} } e^{-\frac{1}{2}\left(\frac{x-\mu}{\sigma}\right)^2}\]

The log-likelihood of a set of observations is:
\begin{align*}
L=&\log\prod_i\frac{1}{\sigma \sqrt{2\pi} } e^{-\frac{1}{2}\left(\frac{x_i-\mu}{\sigma}\right)^2}\\=& \sum_i\left(-\log(\sigma) - \frac{(x_i-\mu)^2}{2\sigma^2  }\right) \\=& -n\log(\sigma)-\frac{1}{2\sigma^2  }\sum_i(x_i-\mu)^2.
\end{align*}

Given the biased sample variance estimator:
\[s^2=\frac{1}{n}\sum_i(\overline{x}-x_i)^2=\overline{x^2}-\overline{x}^2\]
and the formula:
\begin{align*}\sum_i(x_i-\mu)^2=&\sum_i(x_i^2 -2x_i\mu +\mu^2)\\=&\sum_ix_i^2 -2n \mu \overline{x}+n\mu^2
\end{align*},

we can simplify the log-likelihood:

\[L=-n\log(\sigma)-\frac{1}{2\sigma^2  }(ns^2+n\overline{x}^2 -2n \mu \overline{x}+n\mu^2)\]

\[L=-n\left(\log(\sigma)+\frac{s^2}{2\sigma^2  }+\frac{(\overline{x} - \mu)^2}{2\sigma^2}\right)\]

Since the log-likelihood is to be maximized, the loss function is
\(-L\) and we get the equation \eqref{eq:normloss}:

\[\ell = n\left(\log(\sigma)+\frac{s^2}{2\sigma^2  }+\frac{(\overline{x} - \mu)^2}{2\sigma^2}\right)\]

\subsection{Gamma}\label{gamma}
In the case of gamma regression, we have the pdf of the delay given by:

\[f(x)=\frac{\beta^\alpha}{\Gamma(\alpha)} x^{\alpha - 1} e^{-\beta x }\]

\[L=\log\prod_i \frac{\beta^\alpha}{\Gamma(\alpha)} x_i^{\alpha - 1} e^{-\beta x_i }\].

\[L=\sum_i\alpha\log(\beta) + (\alpha-1)\log(x_i)-\beta x_i -\log\Gamma(\alpha)\]
\[L=n (\alpha \log(\beta)+(\alpha-1)\overline{\log(x)}-\beta\overline{x}-\log\Gamma(\alpha)\]

\[\ell = n\left(\log\Gamma(\alpha) + \beta\overline{x} +(1-\alpha)\overline{\log(x)} -\alpha\log(\beta) \right)\]

So for gamma regression we need average delay \(\overline{x}\) and
average log delay \(\overline{\log(x)}\)

\end{document}